\documentclass[aps,pre,showpacs]{revtex4}
\usepackage{graphics}
\usepackage[dvips]{color}
\usepackage{graphicx}
\usepackage{amssymb}
\usepackage{rotating}
\usepackage{epsfig}

\begin{document}
\input epsf.sty


\title{Symmetry based determination of space-time functions in
nonequilibrium growth processes}

\author{Andreas R\"{o}thlein$^1$, Florian Baumann$^{1,2}$, and Michel Pleimling$^{1,3}$}
\affiliation{$^1$Institut f\"{u}r Theoretische Physik I,
Universit\"at Erlangen-N\"{u}rnberg, D -- 91058 Erlangen, Germany\\
$^2$Laboratoire de Physique des Mat\'eriaux (CNRS UMR 7556),
Universit\'e Henri Poincar\'e Nancy I, B.P. 239,
F -- 54506 Vand{\oe}uvre l\`es Nancy Cedex, France\\
$^3$ Department of Physics, Virginia Polytechnic Institute and State University,
Blacksburg, VA 24061-0435, USA}

\begin{abstract}
We study the space-time correlation and response functions in nonequilibrium growth
processes described by linear stochastic Langevin equations. Exploiting exclusively
the existence of space and time dependent symmetries of the noiseless part of these
equations, we derive expressions for the universal scaling functions of two-time
quantities which are found to agree with the exact expressions obtained from the
stochastic equations of motion. The usefulness of the space-time functions is
illustrated through the investigation of two atomistic growth models, the Family
model and the restricted Family model, which are shown to belong to a unique
universality class in 1+1 and in 2+1 space dimensions. This corrects earlier
studies which claimed that in 2+1 dimensions the two models belong to different
universality classes.
\end{abstract}

\maketitle

\section{Introduction}
\label{section1}
Fluctuations are omnipresent when analyzing surfaces and interfaces. These fluctuations 
can be equilibrium fluctuations, as encountered for example when looking at steps on surfaces,
or they can be of nonequilibrium origin as it is the case in various growth processes.
Well known examples of nonequilibrium surface fluctuations are found
in kinetic roughening or nonequilibrium growth processes \cite{Mea93, Hal95, Bar95}
as for example in thin
film growth due to vapor deposition. Interestingly, both equilibrium and nonequilibrium
interface and surface fluctuations 
can be described on a mesoscopic level through rather simple
Langevin equations \cite{Kru97}. In this approach the fast degrees of freedom are modeled by a noise
term, thus yielding stochastic equations of motion for the slow degrees of freedom. In many
instances the physics of dynamical processes is to a large extend captured by linearized
Langevin equations \cite{Kru95} where one distinguishes whether the dynamics is purely diffusive or whether 
mass conservation has to be implemented.

For purely diffusive dynamics (called model A dynamics in critical dynamics \cite{Hoh77,Tau07})
the linear Langevin equation can be written in the following way:
\begin{equation}
  \label{ew1}
  \frac{\partial h(\mathbf{x},t)}{\partial t} = \nu_2 \nabla^2 h(\mathbf{x},t) + \lambda + \eta(\mathbf{x},t)
\end{equation}
where $h(\mathbf{x},t)$ is the value of the macroscopic field $h$ at site $\mathbf{x}$ at time $t$.
In the physical context of fluctuating interfaces and growth processes in $d+1$ spatial dimensions
$h$ is the height field whereas
$\mathbf{x}$ is the lateral position in the underlying $d$ dimensional substrat
lattice. In addition, $\nu_2 > 0$ is the diffusion
constant whereas $\lambda$ is the mean growth velocity (which may of course be zero). 
Finally, the random variable $\eta$ models the noise
due to the fast degrees of freedom. Depending on the physical problem at hand, either Gaussian white noise
or spatially and/or temporally correlated noise is usually considered \cite{Yu94}.

In the context of kinetic roughening and nonequilibrium growth processes Eq.\ (\ref{ew1}) is called the
Edwards-Wilkinson (EW) equation \cite{Edw82}. 
This equation has been used for the description of many dynamical processes,
as for example equilibrium step fluctuations with random attachment/detachment events at the
step edge \cite{Gie01,Dou04,Dou05}. This equation also describes the dynamics of a growing 
surface with a normal incidence of the incoming particles. An obliquely incident particle beam,
however, generates anisotropies which can only be described by a more complex non-linear
Langevin equation \cite{Schm06}.

In growth processes with mass conservation the following linear Langevin equation (with $\nu_4 > 0$)
\begin{equation}
  \label{wv1}
  \frac{ \partial h(\mathbf{x},t)}{\partial t} = -\nu_4 \nabla^4 h(\mathbf{x},t) + \lambda + \eta(\mathbf{x},t)
\end{equation}
has been proposed \cite{Mul63,Wol90}.
This equation is sometimes called the noisy Mullins-Herring (MH) equation.
The noise term again reflects the physics of the investigated system. In the case of equilibrium
fluctuations conserved noise must be considered, leading to the so-called model B dynamics \cite{Hoh77,Tau07}. 
On the other hand, when
studying out-of-equilibrium processes one can again focus on Gaussian white noise or on noise which 
is correlated in space
and/or time \cite{Lam91}. 

The Langevin equation (\ref{wv1}) is used for example to describe
film growth via molecular beam epitaxy \cite{Wol90,Gol91,Das91}, equilibrium fluctuations limited by step
edge diffusion \cite{Dou05,Bon05} or even tumor growth \cite{Bru98,Esc06}.

An important notion in nonequilibrium growth processes is that of dynamical scaling.
Dynamical scaling is nicely illustrated through the behavior of the mean-square width of the 
surface or interface which
for a substrate of linear size $L$ scales as \cite{Fam85}
\begin{equation}
  \label{width2}
W^2(L,t)= L^{2 \zeta} F(t/L^z)
\end{equation}
where $\zeta$ is the roughness exponent and $z$ is the dynamical exponent. For EW we have $z=2$ whereas
for MH $z=4$. The value of $\zeta$ depends on whether correlated or uncorrelated noise is considered.

Langevin equations have the drawback that they do not really mirror the atomistic processes
underlying the fluctuations of the interfaces and surfaces. In order to capture the physics
on the microscopic level one commonly designs simple atomistic models (characterized
by some specific deposition and/or diffusion rules) which are then often
studied numerically. However, it is not always clear what the corresponding Langevin equation is.
Usually, numerical simulations are used in order to extract the exponents $\zeta$ and $z$
(see Eq. (\ref{width2})) which generally permit to relate the microscopic model to one of
the Langevin equations (universality classes). These exponents, however, encode only partly
the information given by a scaling behavior, as the scaling functions, like
$F(y)$ in Eq. (\ref{width2}), are themselves different for different universality classes. 

In this paper we focus on two-point quantities as for example the space-time response and
the space-time correlation functions which also display a dynamical scaling behavior. 
On a more fundamental level
we show, using arguments first given in \cite{Picone04}, that in systems described by the equations (\ref{ew1}) and (\ref{wv1})
the scaling functions of these two-point quantities can be derived by exclusively exploiting the symmetry properties
of the underlying {\it noiseless}, i.e. deterministic, equations. This approach, 
which is based on generalized, space and time dependent,
symmetries of the dynamical system \cite{Henkel01,Henkel02}, has in the past already been applied successfully in the special case
$z=2$ to systems undergoing phase ordering \cite{Henkel03,Henkel04,Picone04,Baumann06a}
and to nonequilibrium phase transitions \cite{Baumann06b}. Here we show that local
space-time symmetries also permit to fix (up to some numerical factors) the scaling
functions of space-time response and correlation functions in cases where $z=4$.
On a more practical level we demonstrate the usefulness of the scaling functions of
these two-point quantities (which depend on two different space-time points ($\mathbf{x}$,$t$) and
($\mathbf{y}$,$s$))
in the characterization of the universality classes of nonequilibrium growth processes.
Whereas in the study of critical systems scaling functions, which are universal and
characterize the different universality classes, are routinely investigated (and this both
at equilibrium \cite{Pel02} and far from equilibrium \cite{Lub04,Cal05}), in
nonequilibrium growth processes the focus usually lies on simple quantities like
for example the exponents $\zeta$ and $z$. There are some notable exceptions where scaling functions
have been discussed (see, for example, \cite{Lam91,Yu94}), but these studies were in general
restricted to one-time quantities.
However, also in nonequilibrium growth processes scaling functions of two-point functions are universal and 
should therefore be very valuable in the determination of the universality class of
a given microscopic model or experimental system.
We illustrate this
by computing through Monte Carlo simulations
the two-point space-time correlation function for two microscopic models which
have been proposed to belong to the same universality class as the Edwards-Wilkinson equation
(\ref{ew1}) with Gaussian white noise \cite{Fam86,Mea87,Liu88,Pal99,Pal03}.

The paper is organized in the following way. In the next Section we discuss the EW and the MH equations
in more detail and introduce the two-point functions. Section III is devoted to the computation of the
exact expressions of the space-time correlation and response functions by Fourier transformation. 
These exact results show {\it inter alia} that the response of the system to the noise
does not depend explicitely on the specific choice of the noise itself.
In Section IV we discuss the space-time
symmetries of the noiseless equations, whereas in Section V we show how these symmetries 
can be used for the
derivation of the scaling functions of two-point functions.
In Section VI we numerically study two microscopic growth models which have been proposed
to belong to the Edwards-Wilkinson universality class.
There has been a recent debate on the universality class of these models which we
resolve by studying the scaling function of the space-time correlation function.
Finally, in Section VII we give our conclusions.
Some technical points are deferred to the Appendices. 

\section{Noise modelization and space-time quantities}
Our main interest in this paper is the investigation of space-time quantities
in systems described by the quite general linear stochastic equations
(\ref{ew1}) and (\ref{wv1}). Setting the mean growth velocity $\lambda$ to zero 
(which can always be achieved by transforming into the co-moving frame) 
both cases can be captured by the single equation
\begin{equation}
\frac{ \partial h(\mathbf{x},t)}{\partial t} = -\nu_{2l} (- \nabla^2)^l h(\mathbf{x},t) + \eta(\mathbf{x},t)
\end{equation}
with $l=1$ (EW) or $l=2$ (MH). As it is well known, these equations of motion can be derived from a free
field theory \cite{Tau07}.

Depending on the physical context, different types of noise may be considered.
For the EW case we shall discuss both Gaussian white noise (EW1)
\begin{equation}
      \label{white_noise}
      \langle \eta (\mathbf{x},t) \rangle = 0, \qquad
      \langle \eta (\mathbf{x},t) \eta (\mathbf{y},s)
      \rangle = 2 D \delta^d (\mathbf{x} - \mathbf{y})
      \delta(t-s)
     \end{equation}
and spatially correlated noise (EW2)
     \begin{equation}
      \label{colored_noise1}
      \langle \eta (\mathbf{x},t) \rangle = 0, \qquad
      \langle \eta (\mathbf{x},t) \eta (\mathbf{y},s)
      \rangle = 2 D \left| \mathbf{x}-\mathbf{y} \right|^{2
      \rho - d}\delta(t-s) 
      \end{equation}
with $0 < \rho < d/2$.
In the past, these two types of noise have been used in the modeling of nonequilibrium growth
processes \cite{Kru97,Kru95}. If, however, one wishes to model thermal equilibrium interface fluctuations,
as for example step fluctuations
rate-limited by evaporation-condensation, one has to consider white noise with the Einstein
relation $D = \nu_2 k_B T$ where $T$ is the temperature and $k_B$ the Boltzmann constant.
For the MH case we also consider the noises (\ref{white_noise}) and (\ref{colored_noise1}),
called MH1 and MH2 in the following.
In this case, however, white noise can only be used in nonequilibrium situations as it breaks the
conservation of mass encoded in the Langevin equation (\ref{wv1}). We shall not consider here
the noisy Mullins-Herring equation with conserved noise
which assures the relaxation towards equilibrium of a system
with conserved dynamics, as this is covered by a forthcoming publication of one of
the authors \cite{Baumann06}.

Two-time quantities have been shown in many circumstances to yield useful insights into the dynamical behavior
of systems far from equilibrium \cite{Cal05}. Of special interest are space and time
dependent functions as for example the space-time
response $R(\mathbf{x},\mathbf{y},t,s)$ or the height-height space-time correlation 
\begin{equation}
\label{correlation}
   C(\mathbf{x},\mathbf{y},t,s) = \langle h(\mathbf{x},t)
   h(\mathbf{y},s) \rangle
\end{equation}
where the brackets indicate an average over the realization of the noise.
The space-time response, defined by
\begin{equation}
\label{response}
   R(\mathbf{x},\mathbf{y},t,s) = \left.\frac{\delta \langle
   h(\mathbf{x},t) \rangle }{\delta j(\mathbf{y},s) }
   \right|_{j=0},
\end{equation}
measures the response of
the interface at time $t$ and position $\mathbf{x}$ to a small perturbation $j(\mathbf{y},s)$ at an earlier
time $s$ and at a different position $\mathbf{y}$ \cite{foot1}. For reasons of causality we have $t>s$. 
At the level of the Langevin equation the perturbation
enters through the addition of $j$ to the right hand side. 
Assuming spatial translation invariance in the directions parallel to the interface, we have
\begin{equation}
\label{trans_inv}
C(\mathbf{x},\mathbf{y},t,s) = C(\mathbf{x}-\mathbf{y},t,s), \qquad R(\mathbf{x},\mathbf{y},t,s) =
R(\mathbf{x}-\mathbf{y},t,s).
\end{equation}
The autocorrelation and autoresponse functions are then defined by
\begin{equation}
C(t,s) := C(\mathbf{0},t,s), \qquad R(t,s) :=
R(\mathbf{0},t,s).
\end{equation}

It is well known that the systems discussed here present a simple dynamical scaling bevahiour
(see Eq. (\ref{width2})). 
For example, for the autoresponse and the autocorrelation functions we have
\begin{equation}
\label{scaling1}
R(t,s) \sim s^{-a-1} f_R(t/s), \qquad C(t,s) \sim s^{-b}
f_C(t/s),
\end{equation}
which defines the nonequilibrium exponents $a$ and $b$. This
terminology is well known from magnetic systems. 
Combining equations
(\ref{width2}), (\ref{correlation}) and (\ref{scaling1}), one
ends up with the scaling relation
$b = -2 \zeta/z$, which relates $b$ to the known exponents $\zeta$ and
$z$. In addition the scaling
functions $f_R$ and $f_C$ define two additional exponents
$\lambda_R$ and $\lambda_C$ by their
asymptotic behavior
\begin{equation}
\label{scaling2}
f_R(y) \stackrel{y \rightarrow \infty}{\sim}
y^{-\lambda_R/z}, \qquad f_C(y) \stackrel{y \rightarrow
\infty}{\sim} y^{-\lambda_C/z}
\end{equation}
where $z$ is again the dynamical exponent introduced in Section I.
Similarly, one obtains for the space-time quantities the following scaling forms:
\begin{equation}
R(\mathbf{x}-\mathbf{y},t,s) \sim s^{-a-1} F_R(\left| \mathbf{x}-\mathbf{y} \right|^z/s,t/s), 
\qquad C(\mathbf{x}-\mathbf{y},t,s) \sim s^{-b} F_C(\left| \mathbf{x}-\mathbf{y} \right|^z/s,t/s).
\end{equation}

\section{Response and correlation functions: exact results}
\label{section3}
This Section is devoted to the computation of space-time quantities
by directly solving the Langevin equations (\ref{ew1}) and (\ref{wv1})
in the physically relevant cases $d=1$ and $d=2$.
These exact results will be used in the following in two different ways.
In Section V we use these expressions in order to check whether our approach,
which exploits exclusively the generalized space-time symmetries of the
deterministic part of the equation of motion, yields the correct results.
In addition, in Section VI we compare these expressions with the numerically determined scaling functions
obtained for two different atomistic models in order
to decide on the universality class of these models.

\subsection{$z=2$: The Edwards-Wilkinson case}
In order to compute the response of the surface/interface to a small perturbation
we add the term $j(\mathbf{x},t)$ to the
right-hand side of the Langevin equation and then go to reciprocal space.
In the EW case the solution of the resulting equation is
(with $d=1$, 2)
\begin{equation}
\label{h_ew_fourier}
\hat{h}(\mathbf{k},t) = e^{-\nu_2 \mathbf{k}^2 t}
\int_0^t d t' e^{\nu_2 \mathbf{k}^2
t'}  (\hat{\eta}(\mathbf{k},t') + \hat{j}(\mathbf{k},t'))
\end{equation}
where we denote by $\hat{h}(\mathbf{k},t)$, $\hat{\eta}(\mathbf{k},t)$
resp. $\hat{j}(\mathbf{k},t)$ the Fourier 
transform of $h(\mathbf{x},t)$, $\eta(\mathbf{x},t)$ resp. $j(\mathbf{x},t)$. We thereby assume flat initial
conditions at time $t=0$, $h(\mathbf{x},0)=0$, i.e. we prepare the system in an out-of-equilibrium
state \cite{footn}.
This preparation 
enables us to study the approach to equilibrium
for the EW1 case with a valid Einstein relation. For the corresponding study of equilibrium dynamical
properties (as encountered in the recent experiments on step fluctuations
\cite{Gie01,Dou04,Dou05}) we have to prepare the system at $t = - \infty$ and replace in 
(\ref{h_ew_fourier}) the
lower integration boundary 0 by $- \infty$. We shall in the following concentrate on 
the out-of-equilibrium situation.

Taking the functional derivative of (\ref{h_ew_fourier}) 
and transforming back to real space yields the result 
\begin{eqnarray}
R(\mathbf{x}-\mathbf{y},t,s) 
& = & \hat{r}_0 \int \frac{d\mathbf{k}}{(2\pi)^d} \, e^{i \mathbf{k} \dot (\mathbf{x}-\mathbf{y})} \,
e^{-\nu_{2} \mathbf{k}^2(t-s)} 
\label{ew_integral_rep} \\
& = & r_0 (t-s)^{-\frac{d}{2}} \exp\left( -
\frac{(\mathbf{x}-\mathbf{y})^2}{4 \nu_2 (t-s)}
\right)
\label{ew_integral_rep2}
\end{eqnarray}
with $r_0 = \frac{1}{(2 \sqrt{\pi \nu_2})^d}$ and $t > s$. 
The exponents $a$ and $\lambda_R$ as well as the scaling function $f_R$ can readily be
obtained from the expression of the autoresponse function (see Equations (\ref{scaling1})
and (\ref{scaling2}))
\begin{equation}
R(t,s) = r_0 (t-s)^{-\frac{d}{2}},
\end{equation}
yielding $a = \frac{d}{2}-1$, $\lambda_R = d$ and $f_R(y) \sim (y-1)^{\frac{d}{2}-1}$.

The expression for the space-time response is completely
independent from choice of the noise term as long as $\langle \hat{\eta}(\mathbf{k},t)
\rangle = 0$. This also holds for the MH case, as 
discussed in the next subsection. In Section V we shall discuss
an alternative way of looking at this fact.

A similar straightforward calculation yields for the space-time correlation the expression
\begin{equation}
\label{res_ew_corr}
C(\mathbf{x}-\mathbf{y},t,s) =  \int_0^t dt' \int_0^s d t'' \int
\frac{d \mathbf{k}}{(2\pi)^{d}} e^{i \mathbf{k} \cdot (\mathbf{x}-\mathbf{y})}
e^{-\nu_2 \mathbf{k}^2 (t+s-t'-t'')} \langle \hat{\eta}(\mathbf{k},t')
\hat{\eta}(-\mathbf{k},t'') \rangle
\end{equation}
where we have exploited the
spatial translation invariance of the noise correlator. For {\it Gaussian white noise} we then obtain
for the space-time correlation
\begin{equation}
\label{full_corr1}
     C(\mathbf{x}-\mathbf{y},t,s)  = 
     c_0 |\mathbf{x}-\mathbf{y}|^{2-d} \left[\Gamma\left(\frac{d}{2}-1,
     \frac{(\mathbf{x}-\mathbf{y})^2}{4\nu_2 (t+s)}\right)-
     \Gamma\left(\frac{d}{2}-1,\frac{(\mathbf{x}-\mathbf{y})^2}
     {4\nu_2 (t-s)}\right)\right]
   \end{equation}
with $c_0 = \frac{D}{2^{d+3} \pi^{3d/2}\nu_2}$.
The autocorrelation function for $d \neq 2$ \cite{Kru97a} is obtained by
   using the known series expansion of the incomplete Gamma-functions $\Gamma$:
   \begin{equation}
   C(t,s) = \frac{2 c_0 (4 \nu_2)^{1-\frac{d}{2}}}{2-d}
   s^{1-\frac{d}{2}} \left[ \left(\frac{t}{s} +
   1\right)^{1-\frac{d}{2}}- \left(\frac{t}{s} -
   1\right)^{1-\frac{d}{2}} \right]
   \end{equation}
from which we find $b = \frac{d}{2}-1 $, $\lambda_C = d $ and 
$f_C(y) \sim 
   (y+1)^{1-\frac{d}{2}} - (y-1)^{1-\frac{d}{2}} $.
For the special case $d=2$ one has to take the logarithmic behavior of the Gamma-functions
into account which yields
\begin{equation}
   C(t,s) = c_0 \ln \frac{t+s}{t-s}.
\end{equation}

For {\it spatially correlated noise} the space-time correlator can only be written as a
series expansion:
\begin{equation}
\label{full_corr2}
C(\mathbf{x}-\mathbf{y},t,s) = \sum\limits_{n=0}^\infty \left( -1 \right)^n a_n^{(d)}(\rho) 
 | \mathbf{x} - \mathbf{y} |^{2n} \left[ (t + s)^{-(2n - 2\rho+ d -2)/2}
- (t - s)^{-(2n - 2\rho+ d -2)/2} \right]
\end{equation}
with $a_n^{(d)}(\rho) = b_n^{(d)} \frac{2^{2\rho} \: D \: \Gamma(\rho) \Gamma(n-\rho + d/2)}{(2 \rho - 2 n -d +2)
\pi^{2 - 3d/2}
\Gamma(d/2-\rho)\nu_2^{n - \rho + d/2} }$ and $b_n^{(1)}=\frac{1}{(2n)!}$, 
$b_n^{(2)}=\frac{1}{((2n)!!)^2}$, 
whereas for the autocorrelator one gets
\begin{equation}
   C(t,s) = a_0^{(d)}(\rho)
   s^{1-\frac{d}{2}+\rho} \left[ \left(\frac{t}{s} +
   1\right)^{1-\frac{d}{2}+\rho}- \left(\frac{t}{s} -
   1\right)^{1-\frac{d}{2}+\rho} \right],
   \end{equation}
   yielding $b = \frac{d}{2}-1+\rho $, $\lambda_C = d - 2\rho $ and $f_C(y) \sim 
   (y+1)^{1-\frac{d}{2}+\rho} - (y-1)^{1-\frac{d}{2}+\rho}$ .

Looking at the expressions (\ref{full_corr1}) and (\ref{full_corr2}), we see that in both cases the
space-time correlation has the following scaling form:
\begin{equation}
\label{C_scaling}
C(\mathbf{x}-\mathbf{y},t,s) = | \mathbf{x} - \mathbf{y} |^\alpha F\left( \frac{(\mathbf{x} - \mathbf{y}
)^2}{t+s},
\frac{(\mathbf{x} - \mathbf{y})^2}{t-s} \right)
\end{equation}
with $\alpha = 2-d$ resp. $2 - d + 2 \rho$ for EW1 resp. EW2. The
scaling function $F$ is then a function of the two scaling variables $\frac{(\mathbf{x} - \mathbf{y} )^2}{t+s}$
and $\frac{(\mathbf{x} - \mathbf{y})^2}{t-s} $.
In the non-equilibrium situation we discuss here the space-time correlation
function is therefore not time translation invariant.
For equilibrium systems which are prepared at $t= - \infty$ time translation invariance
is of course recovered. It is worth noting that the scaling form (\ref{C_scaling}) corrects
the scaling forms given in \cite{Bar95} where only a dependence on 
$\frac{(\mathbf{x} - \mathbf{y})^2}{t-s}$ was predicted far from equilibrium.

\subsection{$z=4$: The Mullins-Herring case}
For the Mullins-Herring case we proceed along the same line as for the Edwards-Wilkinson
case. We here only give the results for one-dimensional interfaces and refer the reader
to the Appendix A for the two-dimensional case.
The solution of the Langevin equation in Fourier space reads in the MH case
\begin{equation}
\label{h_mwv_fourier}
\hat{h}(\mathbf{k},t) = e^{\nu_4 \mathbf{k}^4 t}
\int_0^t d t' e^{-\nu \mathbf{k}^4
t'}  (\hat{\eta}(\mathbf{k},t') + \hat{j}(\mathbf{k},t'))
\end{equation}
which after some algebra yields for one-dimensional interfaces the expression
\begin{eqnarray}
\label{resp_res_wv2}
R(\mathbf{x}-\mathbf{y},t,s) &=& \frac{1}{\pi \nu_4^{1/4}} (\nu_4 (t-s))^{-1/4}\left[
\Gamma\left(\frac{5}{4}\right) {_0F}_2 \left(\frac{1}{2},\frac{3}{4};
\frac{(\mathbf{x}-\mathbf{y})^4}{256 \nu_4 (t-s)}\right)
\right. \nonumber \\ &-& 2
\left.\Gamma\left(\frac{3}{4}\right)
\left( \frac{(\mathbf{x}-\mathbf{y})^4 }{256 (\nu_4
(t-s))}\right)^{1/2} {_0F}_2\left(\frac{5}{4},\frac{3}{2};
\frac{(\mathbf{x}-\mathbf{y})^4}{256 \nu_4 (t-s)}\right) \right]
\end{eqnarray}
for the space-time response. Here the ${_0F}_2$ functions are generalized hypergeometric functions.
It is worth noting that exponentially growing contributions to the 
functions $_0F_2$ just cancel each other, yielding a response function which decreases
for $(\mathbf{x}-\mathbf{y})^4/(t-s) \rightarrow \infty$, as it
should.

The autoresponse function is straightforwardly found (with $t>s$):
\begin{equation}
\label{resp_res_wv3}
R(t,s) = \frac{\Gamma\left(\frac{5}{4}\right)}{\pi \nu_4^{1/4}} (t-s)^{-\frac{1}{4}} 
\end{equation}
which gives us the quantities
$a = - \frac{3}{4}$, $\lambda_R = 1$ and $f_R(y) \sim (y - 1)^{-\frac{1}{4}}$.
One straightforwardly verifies that in any space dimension $d$ one has the relations
$a = \frac{d}{4}-1$, $\lambda_R = d$ and $f_R(y) \sim (y-1)^{-\frac{d}{4}}$.

As for the EW case we remark that the exact expressions 
(\ref{resp_res_wv2}) and (\ref{resp_res_wv3}) are independent of the noise: we obtain the
same results for Gaussian white noise and for spatially correlated noise.
Let us add that we have here only considered perturbations which are not
mass conserving. Whereas this is physically sound for the cases we have in mind here,
one usually considers mass conserving perturbations in the context of critical dynamics
\cite{Tau07,Baumann06}.


Turning to the correlation function, we proceed as for the EW case
and obtain
\begin{equation}
\label{res_mwv_corr}
C(\mathbf{x}-\mathbf{y},t,s) = \int_0^t dt' \int_0^s d
t'' \int \frac{d 
\mathbf{k}}{(2 \pi)^d} e^{i \mathbf{k} \cdot (\mathbf{x}-\mathbf{y})}
e^{-\nu_4 \mathbf{k}^4 (t+s-t'-t'')} \langle \hat{\eta}(\mathbf{k},t')
\hat{\eta}(-\mathbf{k},t'') \rangle .
\end{equation}
For Gaussian white noise this then yields in $1+1$ dimensions the expressions
\begin{equation}
C(\mathbf{x}-\mathbf{y},t,s) =\frac{D}{(2\pi)^2}
\sum\limits_{n=0}^\infty \frac{(-1)^n |\mathbf{x}-\mathbf{y}|^{2n}}{(2n)!
\nu_4^{(2n+1)/4}(3-2n)}
\Gamma\left(\frac{2n+1}{4}\right) \left[(t+s)^{(3-2n)/4}-(t-s)^{(3-2n)/4}\right]
\end{equation}
and
\begin{equation}
C(t,s)=\frac{D\:\Gamma(5/4)}{3\pi2\nu_4^{1/4}}\left[(t+s)^{3/4}-(t-s)^{3/4}\right]
\end{equation}
for the space-time correlation and the autocorrelation functions. Similarly, for
spatial correlated noise we have 
\begin{equation}
C(\mathbf{x}-\mathbf{y},t,s) = \sum_{n=0}^\infty (-1)^n
\tilde{a}_n^{(1)}(\rho)|\mathbf{x}-\mathbf{y}|^{2n}
[(t+s)^{-(2n-2\rho+d-4)/4}-
(t-s)^{-(2n-2\rho+d-4)/4}]
\end{equation}
with $\tilde{a}_n^{(1)}(\rho)=\frac{2^{2\rho}\:D\:\Gamma(\rho)\Gamma((1+2n-2\rho)/4)}{(2n)! (2\rho-2n+3)
\pi^{1/2}\Gamma((1-2\rho)/2)\nu_4^{(1+2n-2\rho)/4}}$, the autocorrelation being given by
the term with $n=0$.

\section{Space-time symmetries of the noiseless equations}
\label{section4}

We have seen in the previous Section that both models under
consideration show dynamical (critical) scaling behavior. For critical
systems it is well know \cite{Tau07,Henkel02}
that the multipoint correlators
$G(\mathbf{x}_1,t_1,\ldots,\mathbf{x}_n,t_n) := \langle
h_1(\mathbf{x}_1,t_1) \ldots h_n(\mathbf{x}_n,t_n) \rangle $  
satisfy a covariance behavior of the kind
\[
  G(b \mathbf{x}_1, b^z t_1, \ldots, b \mathbf{x}_n, b^z t_n) = b^{x_1 + \ldots
  + x_n} G(\mathbf{x}_1,t_1 \ldots, \mathbf{x}_n,t_n)
\]
with a constant rescaling factor $b$ and 
with some numbers $x_i$.
It has been proposed by Henkel \cite{Henkel02} to extend this 
dynamical scaling with $b$ constant to 
rescaling factors which are space and time dependent, i.e. $b
\rightarrow b(\mathbf{x},t)$.  For the special case $z=2$
this can be connected to
so-called Schr\"odinger invariance \cite{Hag72,Nie72}.
In \cite{Henkel01,Henkel02} the concept of
Schr\"odinger invariance has been extended to the case of
an arbitrary value of $z$, but a concrete comparison of the
space-time quantities predicted by this theory with specific,
exactly solvable models with $z \neq 2$ is still lacking. Here we wish to
apply this theory - which is called theory of local 
scale invariance (LSI) - not only 
to the case $z=2$ (i.e. the EW case), but also to the MH case with $z = 4$.

LSI proposes to derive expressions for the scaling functions of two-time
quantities by looking at the space-time symmetries of the {\it noiseless} 
equations of motion. In our case the deterministic equations for $z=2$ resp.
$z=4$ are obtained by dropping the noise term on the right hand side 
of Eq. (\ref{ew1})
resp. (\ref{wv1}). The case $z=2$ then yields the free 
diffusion equation or, equivalently, when going to complex times, the free
Schr\"odinger equation. The maximal kinematic group of space-time symmetries
which leave the free Schr\"odinger equation invariant is the so-called 
Schr\"odinger group \cite{Lie,Kastrup,Hag72,Nie72}, the elements of which transform space and time
in the following way:
\begin{equation}
\mathbf{x} \rightarrow \mathbf{x}' = \frac{\mathcal{R}
\mathbf{x} + \mathbf{v} t + \mathbf{a}}{\gamma t + \delta},
\quad t \rightarrow t' = \frac{\alpha t + \beta}{\gamma t +
\delta},\quad \alpha \delta - \beta \gamma =
1
\end{equation}
where $\mathcal{R}$ is a rotation matrix, whereas
$\mathbf{v},\mathbf{a}$ and $\alpha,\beta,\gamma,\delta$ are
real parameters. We write for this also $(\mathbf{x}',t') =
g(\mathbf{x},t)$ and denote the inverse transformation by
$g^{-1}(\mathbf{x}',t')$. Under the action of these group elements
solutions $\Psi$ of the Schr{\"o}dinger equation transform as
\begin{equation}
  \label{quasiprimary} 
  \Psi \rightarrow \Psi' = f_g(g^{-1}(\mathbf{x},t))
  \Psi(g^{-1}(\mathbf{x},t))  
\end{equation}
where the companion function $f_g$ is known explicitly
\cite{Nie72}. The
generators of the Schr\"odinger group, which can be
considered as infinitesimal version of these
transformations, form a Lie-Algebra.

In \cite{Henkel02} generators of space-time transformations
have been constructed which act as dynamical symmetries on
a more general deterministic equation, namely on
\begin{equation}
\label{symm_operator}
\left[ -\lambda\partial_t +
\frac{1}{z^2} \partial_r^z \right] \Psi =0.
\end{equation}
where $z > 0$ is a real number.
It is easy to see that Equation (\ref{symm_operator}) is equivalent to
(\ref{ew1}) for $z = 2$ resp. to
(\ref{wv1}) for $z = 4$ when setting
$\nu_2 = (4 \lambda)^{-1}$ resp. $\nu_4 = -
(16 \lambda)^{-1}$.
The generators of the corresponding Lie algebra are explicitely given by (for $d = 1$) \cite{Henkel02}
\begin{eqnarray}
\label{generators:xm1}
X_{-1} & = & - \partial_t \\
Y_{-\theta} & = & -\partial_r \\
X_0 &= & -t \partial_t - \frac{1}{z} r \partial_r -
\frac{x}{z} \\
X_1 &= & - t^2 \partial_t - \frac{2}{z} t r \partial_r
- \frac{2 x}{z} t - \lambda r^2 
\partial_r^{2-z}\nonumber \\ && - 2 \gamma_1 (2-z) r
\partial_r^{1-z} - \gamma_1 (2-z)(1-z)
\partial_r^{-z} \\
\label{generators:y2}
Y_{-1/z+1} &=& -t \partial_r - \lambda z r
\partial_r^{2-z} - \gamma_1 z (2-z)
\partial_r^{1-z}
\end{eqnarray}
Operators like
$\partial_r^{1-z}$ are so-called fractional
derivatives, which we recall in the Appendix B. 
Hereby the quantities $x$ and $\gamma_1$ are related by
\begin{equation}
\label{x}
x = \frac{z-1}{2} +
\frac{\gamma_1}{\lambda} (2- z).
\end{equation}


As shown in \cite{Henkel02} these space-time symmetries (\ref{generators:xm1})-(\ref{generators:y2})
can be used to fix
the form of the two-time response function completely. Using all 
generators and writing $\mathbf{r} = \mathbf{x}-\mathbf{y}$, 
one obtains 
\begin{equation}
\label{two_point}
R_0(\mathbf{r},t,s)
= (t-s)^{-2 x/z}
\phi \left(\frac{|\mathbf{r}|}{(t-s)^{1/z}}\right)
\end{equation}
where the index 0 indicates that this is the result for the noise-free theory.
The scaling function $\phi(u)$ satisfies the 
fractional differential equation \cite{Henkel02}
\begin{equation}
\label{frac_diff_equ}
\left(\partial_u + z\lambda u \partial_u^{2-z} + 2
z (2-z) \gamma_1 \partial_u^{1-z} \right)
\phi(u) = 0.
\end{equation}
We have to stress that the scaling function given in \cite{Henkel02} is not 
the most general solution of this equation. In Appendix C we derive this most general
solution for any rational $z$. As shown in the next Section it is this solution which permits us 
to derive the exact expressions for the space-time response and correlation functions
in the MH case by exploiting exclusively the space-time symmetries of the 
deterministic equation (\ref{wv1}).
For $z = 2$ we recover the known result \cite{Henkel02}
\begin{equation}
\label{twopoint_z2}
\phi(u) = \phi_0 \exp \left(-\lambda u^2
\right)
\end{equation}
where the numerical factor $\phi_0$ is not fixed by the theory. For the case $z=4$ and $d=1$
our new result is (see Appendix C for the expression obtained in two space dimensions)
\begin{eqnarray}
\label{twopoint_z4}
\phi(u) &=& \tilde{c}_0 \left( -\frac{\lambda}{16} u^4
\right)^{1/4} \, {_0F_2}\left(\frac{3}{4},\frac{5}{4},
-\frac{\lambda}{16} u^4\right) \nonumber \\ &+&
\tilde{c}_1 \left( -\frac{\lambda}{16} u^4 \right)^{1/2}
\, {_0F_2}\left(\frac{5}{4},\frac{3}{2},
-\frac{\lambda}{16} u^4\right)   + \tilde{c}_2
\,\, {_0F_2}\left(\frac{1}{2},\frac{3}{4},-\frac{\lambda}{16}
u^4\right)
\end{eqnarray}
with some constants $\tilde{c}_0$, $\tilde{c}_1$ and $\tilde{c}_2$. Here we have used the
fact that $x = \frac{d}{2}$ in a free field theory as is easily obtained from a
dimensional analysis (see also the next Section).
The coefficients $\tilde{c}_0$, $\tilde{c}_1$ and $\tilde{c}_2$ have
to be arranged in such a way that $\phi(u)$ vanishes 
for $u \rightarrow \infty$. Analysing the
leadings terms \cite{Wright1,Wright2} one realizes 
that the condition
\begin{equation}
\label{cond_asymptotics}
\Gamma \left(\frac{5}{4}\right)
\Gamma\left(\frac{3}{4}\right) \tilde{c}_0 + 
\Gamma\left(\frac{5}{4}\right) \Gamma
\left(\frac{3}{2}\right) \tilde{c}_1 + 
\Gamma\left(\frac{3}{4}\right) \Gamma\left(\frac{1}{2}\right) \tilde{c}_2 = 0
\end{equation}
provides exactly this, as it cancels all
exponentially growing terms. 

Let us close this Section by noting that in the derivation of expression (\ref{two_point})
we exploited the fact that the exact responses of both the EW and MH case are time
translation invariant. Often when discussing out-of-equilibrium systems this is not the case
and one has to consider the sub-algebra where the generator $X_{-1}$, responsible for
time translation invariance, is omitted
\cite{Henkel01,Henkel02,Henkel03,Henkel06,Picone04}. 
This then yields for the autoresponse the expression
$R_0(t,s) = r_0 s^{-1-a}(\frac{t}{s})^{1+a'-\lambda_R/z}
(\frac{t}{s}-1)^{-1-a'}$, where the parameters $a$ and $a'$ 
have to be determined by comparing with known results. When
setting $a = a' = \lambda_R/z -1$, one recovers our result. 

\section{Determination of response and correlation functions from space-time symmetries}
\label{section5}

In order to use the symmetry considerations of the last Section,
we have to adopt the standard field theoretical setup for
the description of Langevin equations
\cite{Mar73,Jan76,Picone04}. 
Apart from the field $h(\mathbf{x},t)$ we consider the 
so-called response field $\tilde{h}(\mathbf{x},t)$ which leads to the action
\begin{eqnarray}
\label{action}
S[h,\tilde{h}] & =& \int d u \, d \mathbf{R} \left[ \tilde{h}
\left( \partial_u + \nu_{2l} (- \nabla^2)^l \right) h \right] \nonumber \\ 
& & + \frac{1}{2} \int d u \, d \mathbf{R}\,  d u'\,  d
\mathbf{R}' \tilde{h}(\mathbf{R},u) \langle
\eta(\mathbf{R},u) \eta(\mathbf{R}',u') \rangle
\tilde{h}(\mathbf{R}',u')
\end{eqnarray}
where $l=1 $ for the EW case and $l = 2$ for
the MH case. The temporal integration is from
$0$ to $\infty$ whereas the spatial integration is over the
whole space. The following reduction to the
noise-free theory is a well known procedure \cite{Picone04} but we recall the
most important steps in order to establish notations.

Varying the action yields the
equation of motion for the fields $h$ and $\tilde{h}$
\begin{eqnarray}
\frac{\partial h(\mathbf{x},t)}{\partial t} &=& - \nu_{2l} (- \nabla^2)^l
h(\mathbf{x},t)  - \int du \, d\mathbf{R} \,
\tilde{h}(\mathbf{R},u)  \langle
\eta(\mathbf{R},u) \eta(\mathbf{x},t) \rangle \\
\frac{\partial \tilde{h}(\mathbf{x},t)}{\partial t} & = &  \nu_{2l}
(- \nabla^2)^l \tilde{h}(\mathbf{x},t)  
\end{eqnarray}
As expected, one recovers for the height $h(\mathbf{x},t)$ the Equations (\ref{ew1})
and (\ref{wv1}) by identifying the noise with the
term $- \int du \, d\mathbf{R} \,
\tilde{h}(\mathbf{R},u)  \langle
\eta(\mathbf{R},u) \eta(\mathbf{x},t) \rangle$.  

One can now proceed by looking at the multi-point functions which
are defined in the usual way via functional integrals:
\begin{eqnarray}
\label{n_point1}
 \left \langle \prod_{i=1}^n h(\mathbf{x}_i,t_i) \prod_{j=n+1}^m
\tilde{h}(\mathbf{x}_{j},t_j)
\right \rangle  := 
\int \mathcal{D}[h] \mathcal{D}[\tilde{h}] 
\prod_{i=1}^n h(\mathbf{x}_i,t_i) \prod_{j=n+1}^m 
\tilde{h}(\mathbf{x}_{j},t_j)  \exp(-S[h,\tilde{h}]). 
\end{eqnarray}
Within this formalism the space-time response (\ref{response}) is given by
\begin{equation}
R(\mathbf{x},\mathbf{y},t,s) = \langle h(\mathbf{x},t)
\tilde{h}(\mathbf{y},s) \rangle.
\end{equation}

In order to proceed one splits up the action in the same way as done in
\cite{Picone04}, that is as
\begin{equation}
 S[h,\tilde{h}] = S_0[h,\tilde{h}] + S_{th}[h,\tilde{h}]
\end{equation}
with the deterministic part
\begin{equation}
S_0[h,\tilde{h}] = \int d u \, d\mathbf{R} \left[ \tilde{h}(\mathbf{R},u)
\left( \partial_u + \nu_{2l} (- \nabla^2)^l \right) h(\mathbf{R},u) \right] 
\end{equation}
and the noise part
\begin{equation}
S_{th}[h,\tilde{h}] =  \frac{1}{2} \int d u \, d\mathbf{R} \, d u' \, d
\mathbf{R}' \, \tilde{h}(\mathbf{R},u) \langle
\eta(\mathbf{R},u) \eta(\mathbf{R}',u') \rangle
\tilde{h}(\mathbf{R}',u') 
\end{equation}
We call the theory exclusively described by $S_0$ noise-free and
denote averages with respect to this theory with $\langle
\ldots \rangle_0$. The $n$-point functions of the full
theory can then be written as
\begin{equation}
\label{n_point3}
 \left \langle \prod_{i=1}^n h(\mathbf{x}_i,t_i) \prod_{j=n+1}^m
\tilde{h}(\mathbf{x}_{j},t_j)\right \rangle  
=  \left \langle \prod_{i=1}^n h(\mathbf{x}_i,t_i)
\prod_{j=n+1}^m \tilde{h}(\mathbf{x}_{j},t_j)
\exp(-S_{th}[h,\tilde{h}]) \right\rangle_0.
\end{equation}

It is easy to see that the noise-free
theory has a Gaussian structure both for the EW and the MH model. Introducing the
two-component field $\Psi =\left({h \atop \tilde{h}}\right)$
one can write the exponential
$\exp(-S_0[h,\tilde{h}])$ as $\exp(-\int d u \,
d \mathbf{r} \, d u' \, d \mathbf{r}' \, \Psi^t \mathbf{A} \Psi)$
with
\begin{equation}
\mathbf{A} =\frac{1}{2} \left( \begin{array}{cc}
                   0 & \delta(u-u')
		   \delta(\mathbf{r}-\mathbf{r}') ((
		   - \nabla^2)^l - \partial_u)  \\
		   \delta(u-u') \delta(\mathbf{r}-\mathbf{r}') ((- \nabla^2)^l + \partial_u) & 0 
                   \end{array} \right).
\end{equation}

{}From this one deduces two important facts 
which we will need in the sequel.
Firstly, one has
     \begin{equation}
         \label{selection_rule}
	 \langle \underbrace{h \ldots h}_n
	 \underbrace{\tilde{h} \ldots \tilde{h}}_m \rangle_0 =
	 0
     \end{equation}
     unless $n = m$, which is due to the antidiagonal
     structure of $\mathbf{A}$ 
     (see for instance
     \cite{Tau07}, chapter 4) 
     For $z = 2$ Eq. (\ref{selection_rule}) 
     coincides with the Bargmann superselection rule \cite{Bar54}.
Secondly, Wick's theorem holds. With this it follows that
one can write 
the four-point function as
    \begin{eqnarray}
     \label{wick_fourpoint}
     \langle
     h(\mathbf{x},t)h(\mathbf{y},s)\tilde{h}(\mathbf{R},u)
     \tilde{h}(\mathbf{R},u') \rangle_0& =& \langle
     h(\mathbf{x},t) \tilde{h}(\mathbf{R},u) \rangle_0 \langle
     h(\mathbf{y},s) \tilde{h}(\mathbf{R}',u') \rangle_0 \\
     &+& \langle
     h(\mathbf{y},s) \tilde{h}(\mathbf{R},u) \rangle_0\langle
     h(\mathbf{x},t) \tilde{h}(\mathbf{R}',u') \rangle_0
     \nonumber
    \end{eqnarray}
    where we have used Eq. (\ref{selection_rule}). 

Now one can calculate the quantities of interest, namely the space-time
response and correlation functions.  For this one
develops the exponential in (\ref{n_point3}) in a power
series. One remarks immediately that due to the
selection rule (\ref{selection_rule}) one has
\begin{equation}
\label{response_full}
R(\mathbf{x},\mathbf{y},t,s) =
R_0(\mathbf{x},\mathbf{y},t,s) := \langle h(\mathbf{x},t)
\tilde{h}(\mathbf{y},s) \rangle_0,
\end{equation}
i.e. the linear response function of the full theory is {\it equal}
to the noise-less linear response function. It follows that for any realization
of the noise one gets the same expression for the response function, in agreement
with the exact results derived in Section III. It is also worth noting that within
a free field theory non-linear responses vanish due to the
same superselection rule.
Things are of course more tricky for a field theory which is not free as here the
noise can contribute to the reponse. Whether this is the case
depends on the concrete form of the interaction.

By expanding the exponential in (\ref{n_point3}) we obtain in a similar way
that the space-time correlation function is given by the expression
\begin{equation}
\label{correlation_full}
C(\mathbf{x},\mathbf{y},t,s) = \int du  \,d\mathbf{R} \, du' \,
d\mathbf{R}' \, \langle h(\mathbf{x},t) h(\mathbf{y},s) 
\tilde{h}(\mathbf{R},u) \tilde{h}(\mathbf{R}',u') \rangle_0
\langle\eta(\mathbf{R},u)
\eta(\mathbf{R}',u') \rangle.
\end{equation}
Using Wick's theorem we can
replace the four-point function by two-point functions
(see Eq. (\ref{wick_fourpoint})) and obtain
\begin{equation}
\label{correlation_full2}
C(\mathbf{x},\mathbf{y},t,s) = 2 \int du \, d\mathbf{R} \, d u' \,
d\mathbf{R}' \, \langle \eta(\mathbf{R},u) \eta(\mathbf{R}',u') \rangle
\langle h(\mathbf{x},t) \tilde{h}(\mathbf{R},u) \rangle_0
\langle h(\mathbf{y},s) \tilde{h}(\mathbf{R}',u') \rangle_0.
\end{equation}

Inspection of Eqs. (\ref{response_full}) and (\ref{correlation_full2}) reveals
that the only remaining undetermined quantity is the two-point function 
$\langle h(\mathbf{x},t)
\tilde{h}(\mathbf{y},s) \rangle_0$. However, as duscussed in the previous Section,
this two-point function is fully determined by the space-time symmetries of
the deterministic equation of motion. It remains to show that the insertion
of this two-point function into Eqs. (\ref{response_full}) and (\ref{correlation_full2})
indeed yields the exact expressions for the space-time quantities both for the
EW and for the MH case.

\subsection{$z=2$: The Edwards-Wilkinson case}
This case with Gaussian white noise has already been discussed in \cite{Picone04}
in the context of phase ordering kinetics and of critical dynamics. The response function
can be read off from the Equation (\ref{two_point}) after inserting the scaling
function (\ref{twopoint_z2}). Recalling that for the EW case we have $z=2$, $x= \frac{d}{2}$,
and $\nu_2 = (4 \lambda)^{-1}$ we readily obtain the exact result (\ref{ew_integral_rep2}). 
The only quantity left free by the theory is the numerical prefactor $\phi_0$.

The space-time correlation function is obtained by inserting the same two-point
function into Eq. (\ref{correlation_full2}). This is most easily seen by using the
integral representation (\ref{ew_integral_rep}) of the two-point function which yields
after  interchanging the order of integration:
\begin{equation}
C(\mathbf{x},\mathbf{y},t,s) = c_0 \int\limits_0^t dt' \int\limits_0^s dt'' \int \frac{d\mathbf{k}}{(2
\pi)^d}e^{i \mathbf{k}\cdot (\mathbf{x}-\mathbf{y})} 
e^{-\nu_2 \mathbf{k}^2 (t-t')}e^{-\nu_2 \mathbf{k}^2
(s-t'')}\langle \hat{\eta}(\mathbf{k},t')
\hat{\eta}(-\mathbf{k},t'') \rangle
\end{equation}
which is exactly the same expression
(up to the undetermined constant $c_0$) as (\ref{res_ew_corr}), and this for any choice of the
noise correlator $\langle \hat{\eta}(\mathbf{k},t')
\hat{\eta}(-\mathbf{k},t'') \rangle$. It then immediately follows that we recover
the exact results for $C$ given in Section IIIA.

\subsection{$z=4$: The Mullins-Herring case}
For the MH case we proceed along the same lines as for the EW case. 
%
Let us start with the one-dimensional case $d=1$.
As already seen for the EW model, the response function is just the response function
of the noise-free theory whose scaling function is given by (\ref{twopoint_z4})
together with the condition (\ref{cond_asymptotics}) needed for a  
response which vanishes for $u \longrightarrow \infty$. In order to proceed further we remark that
in a powers series expansion of (\ref{twopoint_z4}) odd powers of the scaling variable $u$
only enter through the term with coefficient $\tilde{c}_0$. However, odd powers of
the scaling variable are absent in the exact result (\ref{resp_res_wv2}), so we have to set
$\tilde{c}_0=0$. From Eq. (\ref{cond_asymptotics}) it then follows that 
\begin{equation}
\label{choice_constants}
\Gamma\left( \frac{5}{4} \right)\tilde{c}_1 = -
2 \, \Gamma\left( \frac{3}{4}\right) \tilde{c}_2.
\end{equation}
Recalling that for the MH case $z=4$, $x=\frac{d}{2}$, and $\nu_4 = -
(16 (\lambda)^{-1}$, it is now easy to check that the proposed scaling
function together with (\ref{choice_constants}) indeed yields the exact result
(\ref{resp_res_wv2}) up to the normalization constant $\tilde{c}_2$.
This so determined two-point function can then be inserted into  
the correlation function (\ref{correlation_full2}), yielding the exact
result (\ref{res_mwv_corr}). This is again most easily seen by using
the integral representation
\begin{equation}
R(\mathbf{x}-\mathbf{y},t,s)
 = \hat{r}_0 \int \frac{d\mathbf{k}}{(2\pi)^d} \, e^{i \mathbf{k} \dot (\mathbf{x}-\mathbf{y})} \,
e^{-\nu_{4} \mathbf{k}^4(t-s)}
\end{equation}
of the response function (\ref{resp_res_wv2}).

In two dimensions we have to replace the expression (\ref{twopoint_z4}) by 
(\ref{app:twopoint_z4_d2}). It follows that in $d=2$ the scaling
function obtained from LSI only contains two parameters which are furthermore
related through the condition
(\ref{app:choice2}). We are therefore left with a single undetermined parameter
which only appears as a numerical prefactor, similar to the EW case.
Inserting the resulting scaling function into the expressions
(\ref{response_full}) and (\ref{correlation_full2}) readily yields the exact
results for the space-time quantities in two dimensions.

\section{Microscopic growth models and space-time correlations}
Many theoretical studies of growth processes focus on atomistic models where particles
are deposited on a surface and are then incorporated into the growing
surface following some specific rules which might include local diffusion processes.
Of special interest is the determination of the universality class to which these
models belong. This is usually achieved by computing some universal quantities through
numerical simulations and comparing them to the corresponding quantities 
obtained from continuum growth equations like the EW and the MH equations
discussed in this paper. In a commonly used approach one focuses on the estimation of
the exponents
$z$ and $\zeta$, which govern the behavior of the surface width (\ref{width2}),
through the best data collapse.

In order to show that it is useful to look at two-time quantities 
in nonequilibrium growth processes we discuss
in the following the space-time correlation function in the Family model \cite{Fam86} and
in a variant of this model \cite{Pal99}. Even so these are very simple models, there is still some
debate on the universality class to which these models belong, especially in 2+1 dimensions.
Whereas earlier numerical studies yielded the value $z=2$ for the dynamical exponent
in the 2+1-dimensional Family model \cite{Fam86,Mea87,Liu88,Rei02}, in agreement with the EW universality class
with Gaussian white noise,
Pal et al. \cite{Pal99,Pal03} in their study
obtained a value $z \approx 1.65$, pointing to a different universality class.
In addition they studied a variant of this model (which we call restricted Family model
in the following) for which they recovered $z=2$. These results of Pal et al. are
surprising, especially so as Vvedensky succeeded \cite{Vve03} in deriving in 1+1 dimensions the EW equation 
with Gaussian white noise from
both the Family and the restricted Family model through a coarse-graining procedure.

The Family model is a ballistic deposition model with surface diffusion where a particle is dropped at a
randomly chosen  surface site. Instead of fixing itself at this site, the particle first explores 
the local environment (usually
one restricts this exploration to the nearest neighbors) and fixes itself at the lattice site
with the lowest height. When two or more lattice sites other than the originally selected site 
have the same lowest height, one of these sites is selected randomly.
In case the originally chosen lattice site is among the sites with the lowest height, the particle
remains at this site. In the restricted version of this model, introduced in \cite{Pal99}, the particle
only moves to a site of lowest height when it is unique. This change has the effect that the moving of
the particle only contributes deterministically to the surface shape.

We have simulated these two models both in 1+1 and 2+1 dimensions. For the 1+1 dimensional models all
previous studies agree that $z=2$ and that both models belong to the one-dimensional EW universality class
with Gaussian white noise.
Our main interest here is the height-height space-time correlation function $C(\mathbf{x}
- \mathbf{y},t,s)$. From the 
exact results presented in the first part of the paper we conclude that this two-time quantity should
only depend on the two scaling variables $t/s$ and $r^2/s$ where $r=| \mathbf{x}-\mathbf{y} |$.
In Figure 1 we test this expected scaling behavior in the 1+1 dimensional Family model. In Figure 1a
we fix $t/s$ und plot the correlation function as a function of $r^2/s$, whereas in Figure 1b $r^2/s$
is fixed and C is plotted vs $t/s$. Lattices with 12800 sites have been simulated and the data shown
result from averaging over 1000 runs with different random numbers.
The curves obtained for different values of the waiting time $s$
collapse on a common master curve when multiplying $C$ with $s^{-1/2}$. In addition, these
master curves nicely agree with the expression (\ref{full_corr1}) obtained from the EW equation
with uncorrelated white noise, once the nonuniversal constants $D$ and $\nu_2$ have been determined
\cite{footnote2}.
A similar good agreement is obtained for the restricted Family model.
We list our estimates for $D$ and $\nu_2$ for both one-dimensional models in Table \ref{Table:1}. 
It follows from this table that
$D$ and $\nu_2$ slightly differ in both models.  In addition, $D/\nu_2$ is 
slightly larger for the restricted model, even so the error bars are overlapping.

\begin{figure}[t]
\centerline{\psfig{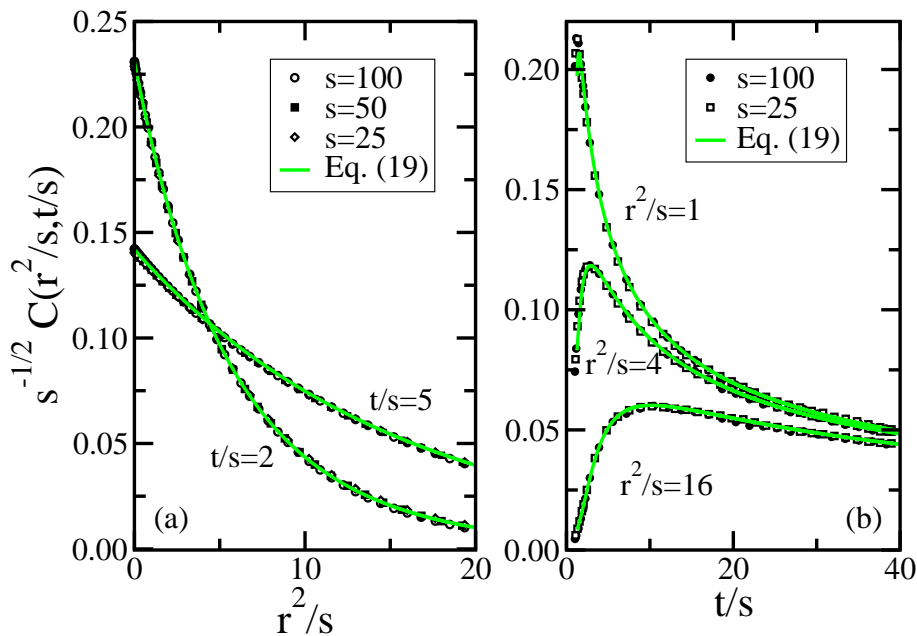}}
\caption{Dynamical scaling of the space-time correlation function
$C(r^2/s,t/s)$ for Family model in 1+1 dimensions with different
values of the waiting time $s$: (a)  $C$ vs $r^2/s$ for some fixed values of $t/s$,
(b) $C$ vs $t/s$ for some fixed values of $r^2/s$. The green curves are obtained
from the exact result (\ref{full_corr1}) derived from the continuum EW equations
with uncorrelated Gaussian white noise. Numerical error bars are smaller than the
sizes of the symbols.}
\end{figure}

It is worthwhile noting that the continuum description is not expected to completely describe
the lattice models for small values of $r^2/s$, as here the discrete nature of the lattice can
not be neglected any more. Deviations are indeed observed in Figure 1b for $r^2/s=1$ and small values
of $t/s$. These lattice effects are expected to get stronger when the dimensionality of the
system increases.

\begin{table}
\caption{\label{Table:1} Estimates for the nonuniversal constants $D$ and $\nu_2$.}
\begin{tabular}{|c|c|c|}
\hline
 & $D$ & $\nu_2$\\
\hline
$d=1$ & 8.85(4) & 1.260(6) \\
\hline
$d=1$ restricted & 9.25(5) & 1.312(7)\\
\hline
$d=2$ & 83(2) & 1.49(5)\\
\hline
$d=2$ restricted & 271(5) & 2.63(6)\\
\hline
\end{tabular}
\end{table}

After having verified that the computed scaling functions in both versions of the 1+1 dimensional
Family model agree with the solution of the EW continuum equation, let us now proceed to the
more controversial 2+1 dimensional case. In Figure 2 we display the space-time correlation
computed for the original Family model in 2+1 dimensions. Again, in the left panel we fix
$t/s$, whereas in the right panel $r^2/s$ is kept constant. The data shown here have been obtained
for lattices with $300 \times 300$ sites with 5000 runs for every waiting time. 
We carefully checked that our nonequilibrium data are
not affected by finite-size effects. Furthermore, we ran different simulations with different random number
generators and obtained the same results within error bars.
We obtain as the main result of these simulations that the scaling function of the space-time
correlation function is in excellent agreement (once the values of the
nonuniversal constants have been determined, see Table 1) with the exact result obtained from solving the
two-dimensional EW equation with uncorrelated white noise. The only discrepancies observed
for small $r^2/s$ and small $t/s$ are qualitativly the same as for the 1+1 case and are
explained by the discrete nature of the lattice. Our results are in accordance with the results
of \cite{Fam86,Mea87,Liu88,Rei02,Vve03} but strongly disagree with those of Pal et al. 
\cite{Pal99,Pal03}. Indeed, a noninteger value of $z$ in a continuum description can not be realized in 
a linear stochastic differential equation and leads to completely different scaling functions as
those obtained from the EW equation.   

\begin{figure}[t]
\centerline{\psfig{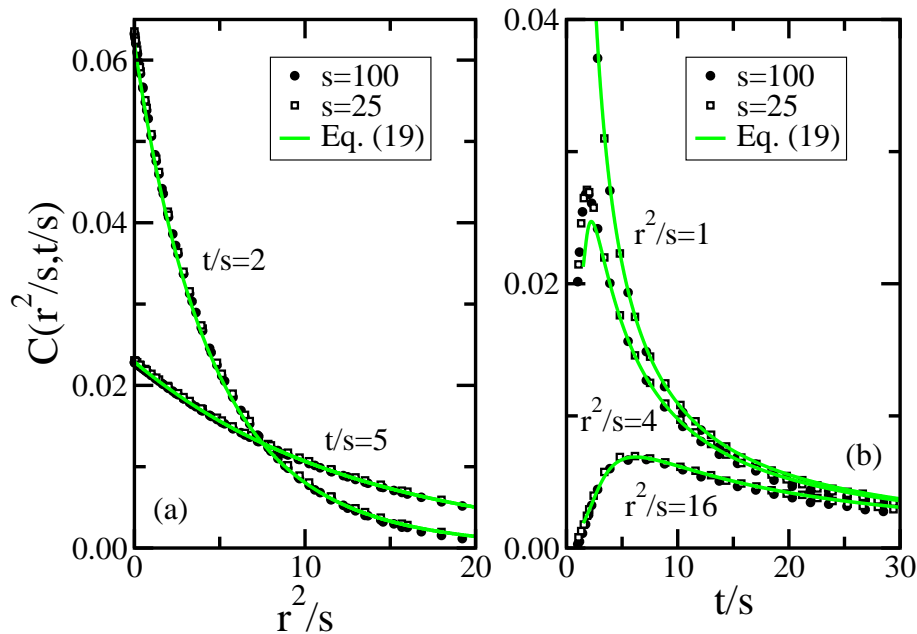}}
\caption{The same as in Figure 1, but now for the Family model in 2+1 dimensions.
Numerical error bars are comparable to the
sizes of the symbols.}
\end{figure}

In Figure 3 (see also Table 1) we show our results for the restricted family model in 
2+1 dimensions. Again, dynamical scaling is observed, and again the data are well described
by the EW scaling functions in the scaling limit. However, the determined values of the nonuniversal
quantities $D$ and $\nu_2$ are markedly different from the values obtained for the original
model. Specifically, the ratio $D/\nu_2$ (which is of the dimension $k_B T$)
is much larger for the restricted model. Identifying $D/\nu_2$ with a (nonequilibrium) temperature,
we can view the processes in the restricted model to take place at a higher 
temperature than in the original model. 
This is in agreement with the observation from Pal et al. \cite{Pal03} that 
the surface is locally rougher in the restricted model, as evidenced by the larger value
of the interface width. In addition, the change in the diffusion rule leads to a nonmonotoneous
behavior of
the correlation function for small $r^2/s$, as shown in the inset of Figure 3a 
for $s=25$ and $t/s=1.04$. Plotting the correlation function in both the (10) and the (11) direction,
we see that correlations between nearest neighbours are suppressed, whereas the autocorrelation,
i.e. the correlation with $r=0$, is strongly enhanced. This behavior can be understood by
recalling that in the restricted model a particle only diffuses to a lower nearest neighbor site
when this site is unique, but otherwise remains on the original site.
If we increase $s$ and $r$, this effect weakens, and it completely vanishes in the scaling
limit of large waiting times and large values of $r^2/s$.

\begin{figure}[t]
\centerline{\psfig{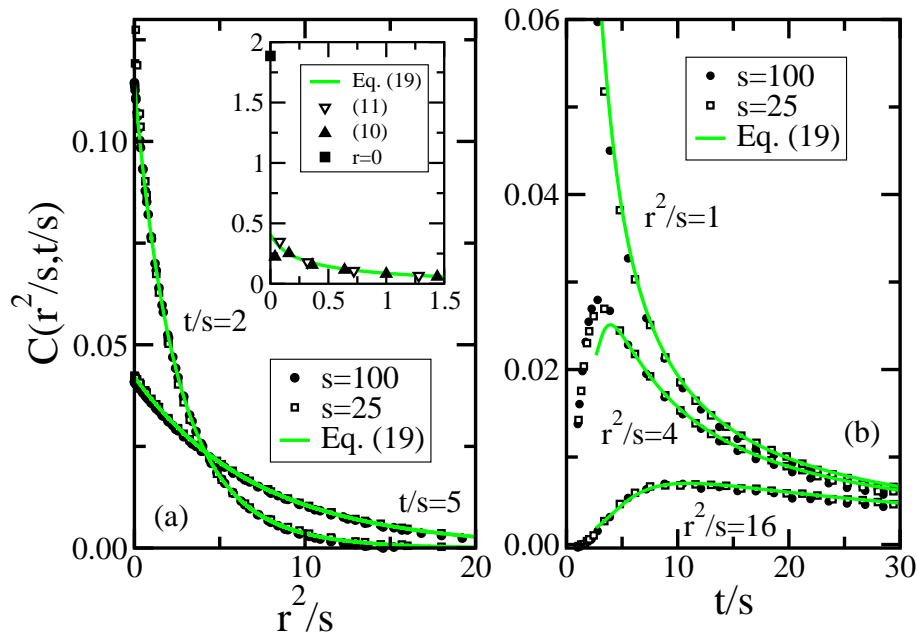}}
\caption{The same as in Figure 2, but now for the restricted Family model in 2+1 dimensions.
The inset in (a) shows the correlation function in the (10) and (11) directions for the
case $s=25$ and $t/s=1.04$. The change of the diffusion rule has a strong impact on the 
autocorrelation with $r=0$ and on the nearest neighbor correlations.
Numerical error bars are comparable to the
sizes of the symbols.}
\end{figure}

{}From our observation that the numerically computed space-time correlation functions of
both microscropic models coincide in the scaling limit with the exact expression from the
EW model we conclude that both the Family model and the restricted Family model belong 
to the EW universality class with uncorrelated noise, and this not only in 1+1 dimensions
but also in 2+1 dimensions.

\section{Conclusions}
\label{section7}
The aim of the present paper is twofold: on the one hand we discuss the usefulness of
universal scaling functions of space-time quantities in characterizing the universality
class of nonequilibrium growth models, on the other hand we demonstrate how a general
symmetry principle allows to derive scaling functions of two-point quantities for equilibrium
and nonequilibrium processes described by linear stochastic Langevin equations.

In the context of nonequilibrium growth processes it is rather uncommon to study universal scaling
functions of two-point quantities in the dynamical scaling limit
in order to determine the universality class to which a given microscopic model
belongs. We have illustrated the usefulness of this approach by comparing the numerically 
obtained space-time correlation functions for two atomistic growth models with the exact
expressions obtained from the corresponding continuum stochastic Langevin equation. This approach has allowed us
to show that both models belong to the same universality class, thus correcting conclusions
obtained in earlier numerical studies.

The study of universal scaling functions of space-time quantities should also be of value
in more complex growth processes which are no more described by linear stochastic differential
equations. Examples include ballistic deposition with an oblique incident particle beam \cite{Schm06}
or growth processes of the KPZ \cite{Kar86} and related \cite{Lai91} universality classes 
where nonlinear effects can no more
be neglected. In addition, the scaling functions studied here can also be measured in experiments
involving nonequilibrium or equilibrium interface fluctuations.
A promising system is given by equilibrium step fluctuations 
\cite{Gie01,Dou04,Dou05,Bon05}, as these are again described by linear Langevin equations.

In addition we have shown that  in nonequilibrium growth processes
scaling functions of out-of-equilibrium quantities
can be derived in a {\it model independent} way by exploiting 
the generalized space-time symmetries
of the {\it noiseless} part of the stochastic equations of motion. 
We have demonstrated explicitly
how to proceed in the case of a rational dynamical exponent $z$, following the general ideas
formulated by Henkel a few years ago \cite{Henkel02}. In the context of nonequilibrium growth
processes, the cases $z=2$ (EW model) and $z=4$ (MH model) are of special interest. The case $z=2$
has already been studied extensively in the past. For the case $z=4$, however, we present to our
knowledge for the first time the derivation of nonequilibrium scaling functions by exploiting
the mentioned symmetry principles. As these scaling functions are found to agree with the exact
expressions derived from the MH equation, we conclude that the postulated space-time symmetries
and the proposed way for constructing the scaling functions
can also be valid for other cases than
merely the case $z=2$.

Let us end this paper by a general remark on the applicability of the concept of local scale
transformations in the context of other out-of-equilibrium processes. The data presented in this
paper for the microscopic growth models nicely show the limitations of the continuum equations
in describing atomistic models. Whereas in the limit of large times and large spatial separations
the numerically computed correlation functions completely agree with the exact results from
the continuum equation, notable deviations are observed for small times and small spatial
separations. These deviations reflect the microscopic details of the models (underlying lattice structure,
diffusion rules etc.) which are not captured by the continuum model. This sheds an interesting
light on an ongoing discussion \cite{Ple05,Hin06,Henkel06,Lip06}
on the applicability of the theory of local scale invariance, which, we 
recall, permits to derive expressions for scaling functions starting from the noiseless part of the
{\it continuum} equations of motion. Clearly, it has to be expected that the so derived
scaling functions can not fully describe microscopic models in the short time and short distance
limit. As it is exactly this limit which is in the centre of the mentioned discussion involving
the theory of local scale invariance, it seems advisable to take any observed deviations in this limit between
numerical data, obtained from simulations of microscopic models, and the theoretically derived scaling
functions {\it cum grano salis}, as these deviations might only reflect the microscopic nature of
the models.

\appendix
\section{Exact results for the Mullins-Herring case in $d=2$}
We here compile the exact results in two space dimensions for the dynamical two-time quantities 
for the Mullins-Herring case with $z=4$. The space-time response can again be expressed
through generalized hypergeometric functions and reads
\begin{eqnarray}
R(\mathbf{x} - \mathbf{y}, t,s)& = &\frac{1}{8\pi\nu_4^{1/2}} \left[ \pi^{1/2} {_0}F_2\left( \frac{1}{2},1,\frac{1}{256\nu_4}
\left( \frac{|\mathbf{x}-\mathbf{y}|}{(t-s)^{1/4}} \right)^4  \right) \right. \nonumber \\
& & \left. -\frac{1}{4\nu_4^{1/2}} 
\left( \frac{|\mathbf{x}-\mathbf{y}|}{(t-s)^{1/4}} \right)^2 {_0}F_2 \left( 1,\frac{3}{2},
\frac{1}{256\nu_4} \left( \frac{|\mathbf{x}-\mathbf{y}|}{(t-s)^{1/4}}\right)^4 \right) \right].
\end{eqnarray}
For the autoresponse one gets
\begin{equation}
R(t,s)=\frac{1}{8\pi^{1/2}\nu_4^{1/2}}\left( t-s \right)^{-1/2}.
\end{equation}
As already noted before, these expressions give the response of the system to the noise itself
and therefore do not depend on the concrete realization of the noise as long as it is nonconserving.

For the space-time correlation function $C(\mathbf{x}-\mathbf{y},t,s)$ and for the autocorrelation
function $C(t,s)$ we obtain the following expression, which depend on the form of the noise:

Gaussian white noise (MH1):
\begin{eqnarray}
C(\mathbf{x}-\mathbf{y},t,s) & = & \frac{D}{32\pi^2} \left[ \left( \sum\limits_{n=0,n\neq 1}^\infty
\frac{(-1)^n |\mathbf{x}-\mathbf{y}|^{2n} \Gamma(\frac{n+1}{2})}{(2n)!\Gamma(1+2n)(\Gamma(\frac{1}{2}
-n))^2\nu_4^{(n+1)/2}(1-n)} \right. \right. \nonumber \\
&& \left. \left. \cdot \left[ (t+s)^{(1-n)/2} - (t-s)^{(1-n)/2} \right] \Bigg) \right. - \frac{|\mathbf{x}-\mathbf{y}|^2}{4
\pi \nu_4} \ln \left( \frac{t+s}{t-s} \right) \right], \\
C(t,s) & = & \frac{D}{32\pi^{5/2}\nu_4^{1/2}}\left[ (t+s)^{1/2} - (t-s)^{1/2} \right].
\end{eqnarray}

Spatially correlated noise (MH2):
\begin{eqnarray}
C(\mathbf{x}-\mathbf{y},t,s)  & =&   \sum\limits_{n=0}^\infty (-1)^n \tilde{a}_n^{(2)}(\rho)
|\mathbf{x}-\mathbf{y}|^{2n} \left[ (t+s)^{-(2n-2\rho-2)/4}-(t-s)^{-(2n-2\rho-2)/4} \right], \\
C(t,s) & = & \tilde{a}_0^{(2)}(\rho) \left[ (t+s)^{(2+2\rho)/4}-(t-s)^{(2+2\rho)/4} \right]
\end{eqnarray}
with $\tilde{a}_n^{(2)}(\rho)=\frac{2^{2\rho-1}\:D\:\Gamma(\rho)\Gamma((2+2n-2\rho)/4)}{2^{2n}(n!)^2 
(2\rho-2n+2)
\Gamma(1-\rho)\nu_4^{(2+2n-2\rho)/4}}$.

\section{On fractional derivatives}

We list here the most important properties of the fractional
derivates as these will be used in Appendix C in the derivation of the
scaling function of the space-time response. We stress the
point that there are several definitions of fractional
derivatives available, which are not equivalent. However, we
need a special type of fractional derivatives. We
simply quote the most important properties as given in
\cite{Henkel02} and refer the reader to this reference for a
more thorough introduction.

$\partial_r^a$ acts on a function $f(r)$ which can be expanded
into the form $f(r) = \sum_{e \in E}^\infty f_e r^{e} +
\sum_{n=0}^\infty F_n \delta^{(n)}(r)$. Here E is the set $E
= \mu \mathbb{N} + \lambda$ with $\mu > 0$ and $\lambda \neq
- (\mu (n+1) + m +1)$, where $n,m \in \mathbb{N}$.
$\delta^{(n)}$ is the $n$-th derivate of the delta
function. $\partial^a_r$ is then defined by
\begin{eqnarray}
   i.) & & \partial_r^a(\alpha f(r) + \beta g(r)) = \alpha
   \partial_r^a f(r) + \beta \partial_r^a g(r) \\
   ii.) &&\partial^a r^e =  \frac{\Gamma(e+1)}{\Gamma(e-a+1)} r^{e-a} +
   \sum_{n=0}^\infty \delta_{a,e+n+1} \Gamma(e+1)
   \delta^{(n)}(r) \\ 
   iii.) & & \partial_r^a \delta^{(n)}(r) =
   \frac{r^{-1-n-a}}{\Gamma(-a-n)} + \sum_{m=0}^\infty
   \delta_{a,m} \delta^{(n+m)}(r)
\end{eqnarray}
Here $\alpha$ and $\beta$ are real constants and $g(r)$ is
another function which can be expanded in the same way as $f(r)$.
The most important properties of these fractional
derivatives are:
\begin{eqnarray}
\label{prop:comm}
\partial^{a+b}_r f(r) & = & \partial_r^a \partial_r^b f(r) =
\partial_r^b \partial_r^a f(r) \\
{[}\partial_r^a,r] f(r) &=& (\partial_r r - r \partial_r) f(r) = a \partial_r^{a-1} f(r) \\
\partial_r^a f(\alpha r) & = & \alpha^a \partial^a_{\alpha
r} f(\alpha r) \\
\partial_{\alpha r}^a f(r) &=& \alpha^{-a} \partial_r^a f(r)
\end{eqnarray}

\section{Derivation of the scaling function}

In this Appendix we outline the derivation of the scaling
function $\phi(u)$ for any rational dynamical exponent $z$, thereby correcting
the incomplete result given in
\cite{Henkel02}. For notational simplicity we do this in one space dimension.
For $z=2$ and $z=4$ we give the results in two space dimensions at the end of this Appendix.

Our starting point is the equation
(\ref{frac_diff_equ}):
\begin{equation}
  \label{appb:diff1}
  \left( \partial_u + \hat{a} u \partial_u^{2-z} + \hat{b}
  \partial_u^{1-z} \right) \phi(u) = 0
\end{equation}
with $\hat{a}= z \lambda$ and $\hat{b} = 2 z (2-
z) \gamma_1$. We write the rational dynamical exponent as $z = N +
\frac{p}{q}$, where $N$ is the largest integer equal or
smaller than $z$. We also assume $\hat{a} \neq 0$ and give the
result for $\hat{a} = 0$ at the end, as it can be derived in
exactly the same way. In a first step we rewrite (\ref{appb:diff1}) as
\begin{equation}
  \label{appb:diff2}
  \left( \partial_u^z + \hat{a} u \partial_u + \hat{b} \right)
  \Psi(u) = 0
\end{equation}
with 
$\Psi(u) = \partial_u^{1-z} \phi(u)$. In doing so we 
have used property (\ref{prop:comm}) of the
fractional derivative. It is in fact this step which
enables us to avoid the negative exponents for the
fractional derivatives in equation (\ref{appb:diff1}), which
are responsible for the incomplete result in \cite{Henkel02}. 
The solution $\Psi(u)$ of this equation yields
then the scaling function $\phi(u)$ through the relation 
\begin{equation}
  \label{appb:diff3}
\phi(u) = \partial_u^{z-1}\Psi(u). 
\end{equation}
Before doing this, it 
is instructive to consider (\ref{appb:diff2}) for the case
$z = 4$. Indeed, Eq. (\ref{appb:diff2}) is then a normal
differential equation of forth order, so that the solution
will be, a priori, a linear combination of four linearly
independent solutions (i.e. it will contain four free
parameters). The method applied in \cite{Henkel02}, however, only yields
one of these linearly independent solutions.

We solve (\ref{appb:diff2}) for
$u>0$ under the additional assumption that the desired solution is nonsingular
for $u \rightarrow 0$. Furthermore, we require that the
scaling function should drop to zero for $u \rightarrow
\infty$.  We then make the ansatz
\begin{equation}
\Psi(u) = \sum_{n=0}^\infty c_n u^{\frac{n}{q} + s}, \qquad
c_0 \neq 0
\end{equation}
and suppose $s > -1$ \cite{footnote}.
This ansatz is introduced into (\ref{appb:diff2}) and yields, because
of $c_0 \neq 0$, the recursion relation
\begin{equation}
  c_{n+\xi} \frac{\Gamma((n+q)/q + s + 1)}{\Gamma(n/q + s +
  1)} + c_n \left( \hat{a}(n/q + s) + \hat{b} \right) = 0
\end{equation}
with $\xi = p + q N$ as well as the relation
\begin{equation}
s = \frac{p}{q} + m, \qquad m \in \mathcal{E},
\end{equation}
where the set $\mathcal{E}$ is given by
\begin{equation}
\mathcal{E} := \left\{ \begin{array}{cc}
-1,0,\ldots, N-1, &  p \neq 0 \\
0,\ldots,N-1, &  p = 0
           \end{array} \right.
\end{equation}
It then also follows that
$c_1 = \ldots = c_{p+q N -1} = 0$.
With this we obtain after some algebra:
\begin{equation}
\Psi(u) = \sum_{m \in \mathcal{E}} c_m \sum_{n = 0}^\infty a_n^{(m)} 
u^{n z + \frac{p}{q} + m}
\end{equation}
with 
\begin{equation}
a_n^{(m)} = \frac{(-\hat{a} z)^n \Gamma 
\left(\frac{p}{q} + 1 + m \right)
\Gamma \left(n + \frac{p/q +m}{z} + \frac{\hat{b}/\hat{a}}{z}\right)}
{\Gamma \left( n z + \frac{p}{q} + +m +1 \right)
\Gamma \left( \frac{p/q + m}{z} + \frac{\hat{b}/\hat{a}}{z} \right)}.
\end{equation}
The $c_m$ are free parameters not fixed by the theory.
Finally, the scaling function $\phi(u)$ is obtained from Eq. (\ref{appb:diff3}):
\begin{equation}
\label{result:sf}
\phi(u) = \sum_{m \in \mathcal{E}} c_m \phi^{(m)}(u)
\end{equation}
with
\begin{equation}
\phi^{(m)} (u)= \sum_{n = 0}^\infty
b_n^{(m)} u^{(n-1) z + \frac{p}{q} + m +1}
\end{equation}
where the coefficients $b_n^{(m)}$ are given by
\begin{equation}
\label{bn}
b_n^{(m)} = \frac{(-\hat{a} z)^n \Gamma \left(\frac{p}{q} + 1
+ m \right)
\Gamma \left(n + \frac{p/q +m}{z} + \frac{\hat{b}/\hat{a}}{z}
\right)}{\Gamma \left( (n-1) z + \frac{p}{q} +
m +2 \right) \Gamma \left( \frac{p/q + m}{z} +
\frac{\hat{b}/\hat{a}}{z} \right)}.
\end{equation}

We remark that
our final result (\ref{result:sf}) is indeed regular for $u
\rightarrow 0$, as $b_0^{(m)} = 0$ for $m = -1,\ldots,N-2$. This is
readily seen by recalling that $\Gamma(l)=\infty$ for $l \in 
- \mathbb{N}_0$. As already mentioned in \cite{Henkel02},
the radius of convergence is infinite for $z > 1$.

We also note that the number of free parameters is a priori
equal to $N$ if $z \in \mathbb{N}$ and $N+1$ else. However,
there might be cases where some of the independent
solutions $\phi^{(m)}(u)$ vanish. 

For completeness let us
also quote the result for $\hat{a} = 0$. In this case
\begin{equation}
b_n^{(m)} = (-\hat{b})^n \frac{\Gamma \left(\frac{p}{q} + 1
+ m \right)
\Gamma \left(n z + \frac{p}{q} +m + 1 \right)}{\Gamma \left( (n+1) z
+ 1 \right) \Gamma \left((n-1) z + \frac{p}{q}+m+2 \right)}.
\end{equation}

The expressions (\ref{twopoint_z2}) and (\ref{twopoint_z4})
for $\phi(u)$ can be obtaind from (\ref{result:sf}) by
setting $z = 2$ (i.e.
$p = 0$ and $N = 2$) or $z = 4$ (i.e. $p = 0$ and $N =
4$) respectively. 
For the EW case it is important to note 
that  $\phi^{(0)}(u)$ vanishes as $b_n^{(0)} = 0$ for every $n$ which immediately
follows from the fact that $\hat{b}= 2z(2-z)\gamma_1 = 0$ for $z=2$.
The remaining solution $\phi^{(1)}(u)$ is then just the
exponential function
\begin{equation}
\label{app:twopoint_z2}
\phi^{(1)}(u) = \exp \left(-\lambda u^2
\right).
\end{equation}
For the MH case it is the solution $\phi^{(2)}$ which vanishes
in the free-field case. Indeed, from Eq. (\ref{x}) we obtain
$\frac{\hat{b}}{4 \hat{a}}
= - \frac{1}{2}$ by recalling that $x = \frac{1}{2}$. It then follows that
the Gamma-function $\Gamma \left( \frac{p/q + m}{z} +
\frac{\hat{b}/\hat{a}}{z} \right)$
always diverges in the denominator of Eq. (\ref{bn}), yielding $\phi^{(2)}(u) = 0$ for
every $u$. We are therefore left with three independent solutions, and after relabelling we
obtain the final
expression (\ref{twopoint_z4}):
\begin{eqnarray}
\phi(u) &=& \tilde{c}_0 \left( -\frac{\lambda}{16} u^4
\right)^{1/4} \, {_0F_2}\left(\frac{3}{4},\frac{5}{4},
-\frac{\lambda}{16} u^4\right) \nonumber \\ &+&
\tilde{c}_1 \left( -\frac{\lambda}{16} u^4 \right)^{1/2}
\, {_0F_2}\left(\frac{5}{4},\frac{3}{2},
-\frac{\lambda}{16} u^4\right)   + \tilde{c}_2
\, {_0F_2}\left(\frac{1}{2},\frac{3}{4},-\frac{\lambda}{16}
u^4\right).
\end{eqnarray}

Let us add that the asymptotic behavior of the generalized hypergeometric functions
${_0F_2}$ is well known \cite{Wright1,Wright2}.
Recalling that the scaling function should vanish for $u \rightarrow \infty$,
we can exploit this known asymptotic behavior in order to derive relations
between the parameters $c_m$. For the case $z=4$ this then yields the condition
(\ref{cond_asymptotics}) given in Section \ref{section4}.

Let us finish this Appendix by quoting the resulting scaling functions $\phi(u)$
in two space dimensions. For the EW case with $z=2$ we get the same expression
(\ref{app:twopoint_z2}) as for the one-dimensional case. For the MH case with $z=4$, our calculations
yield the expression
\begin{equation}
\label{app:twopoint_z4_d2}
\phi(u) = \tilde{c}_1 \left( -\frac{\lambda}{16} u^4 \right)^{1/2}
\, {_0F_2}\left(\frac{3}{2}, \frac{3}{2}, -\frac{\lambda}{16} u^4\right)   + \tilde{c}_2
\, {_0F_2}\left(\frac{1}{2}, 1, -\frac{\lambda}{16} u^4\right)
\end{equation}
with the additional condition
\begin{equation}
\label{app:choice2}
\tilde{c}_1 = - \frac{4}{\sqrt{\pi}} \tilde{c}_2.
\end{equation}

\begin{acknowledgments}
We acknowledge the support by the Deutsche Forschungsgemeinschaft
through grant no. PL 323/2.
This work  was supported by the franco-german binational
programme PROCOPE. The authors are grateful to Malte Henkel
for useful comments and helpful discussions.
\end{acknowledgments}


\begin{thebibliography}{99}
\bibitem{foot1} The response function (\ref{response}) can also be viewed as the response
of the system to the noise itself.
\bibitem{footn} This assumption is also important in
connection with the symmetry based approach presented
in this paper. A nonvanishing intial average of
$h(\mathbf{x},0)$ would lead to modifications.
\bibitem{footnote2} We follow here the procedure used in \cite{Henkel03}
in the context of phase ordering kinetics.
\bibitem{footnote} One can show that this is
justified because of the regularity assumption for $u
\rightarrow 0$. It also follows from the regularity assumption that
a term of the type $\sum_n F_n \delta^{(n)}(u)$ needs not to be included.
\end{thebibliography}
\end{document}